\documentclass[aps,pre,twocolumn,showpacs,groupedaddress]{revtex4}
\usepackage{graphicx,amsmath}
\begin{document}

\title{Viscoplasticity and large-scale chain relaxation in glassy-polymeric strain hardening}
\author{Robert S. Hoy}
\email{robert.hoy@yale.edu}
\author{Corey S. O'Hern}
\affiliation{Department of Mechanical Engineering, Yale University, New Haven, CT 06520-8286}
\affiliation{Department of Physics, Yale University, New Haven, CT 06520-8120}
\date{\today}
\begin{abstract}
A simple theory for glassy polymeric mechanical response that accounts for large scale chain relaxation is presented.
It captures the crossover from perfect-plastic response to Gaussian strain hardening as the degree of polymerization $N$ increases, without invoking entanglements.
By relating hardening to interactions on the scale of monomers and chain segments, we correctly predict its magnitude.
Strain activated relaxation arising from the need to maintain constant chain contour length reduces the characteristic relaxation time by a factor $\sim \dot\epsilon N$ during active deformation at strain rate $\dot\epsilon$.
This prediction is consistent with results from recent experiments and simulations, and we suggest how it may be further tested experimentally.
\end{abstract}
\maketitle

\section{Introduction}

Developing a microscopic, analytic theory of glassy polymeric mechanical response has been a longstanding challenge.
Plasticity in amorphous materials is almost always \textit{viscoplasticity}, i.\ e.\ plasticity with rate dependence.
Many recent studies have focused on plasticity in metallic or colloidal glasses.
Relative to these systems, polymer glasses possess a wider range of characteristic length and time scales, because of the connectivity, uncrossability, and random-walk-like structure of the constituent chains.
These alter the mechanical properties significantly \cite{haward97}; for example, a uniquely polymeric feature of plastic response is massive strain hardening beyond yield.

The relationships between polymeric relaxation times on different spatial scales are fairly well understood for  melts \cite{doi86}, but much less so for glasses. 
Recent experiments \cite{loo00,lee09,lee09b}, simulations \cite{capaldi02,riggleman08,lee09b}, and theories \cite{chen07,chen09} have all shown that local (segment-level) relaxation times in polymer glasses decrease dramatically under active deformation (especially at yield) and increase when deformation is ceased. 
Analysis of this phenomena has focused on stress-assisted thermal activation of the local relaxation processes, but it is likely that other structural relaxation processes at larger scales or of different (e.\ g.\ strain activated) character 
are also important in determining the mechanical response.
Improved understanding of the concomitant scale-dependent relaxation is necessary to better understand polymeric plasticity and material failure.
However, theoretical prediction of large-scale relaxation in deformed polymer glasses is still in its infancy; most treatments evaluate mechanical response ``neglecting the effect of the ongoing structural relaxation during the experiment'' \cite{grassia09}. 

This paper is an attempt to improve on this state of affairs.
We develop a theory that treats large-scale relaxation of uncrosslinked chains during active deformation.
Stress in the postyield regime is assumed to arise from the local plastic rearrangements similar to those which control plastic flow; as strain increases, these increase in rate with the volume over which they are correlated.
Polymeric strain hardening is thus cast as plastic flow in a medium where the effective flow stress increases with large scale chain orientation.
Relaxation of chain orientation is treated here as inherently strain activated and \textit{coherent}, i.\ e.\ cooperative along the chain backbone.

Our theory predicts a continuous crossover from perfect plasticity to ``Gaussian'' (Neohookean) \cite{haward93} strain hardening as the degree of polymerization $N$ increases. 
The latter form is predicted when chains deform affinely on large scales, and corresponds to the limit  where the system is deformed faster than chains can relax.
A key difference from most previous theories is that instead of invoking entanglements, we relate the timescale $\tau$ for large scale chain relaxation to the segmental relaxation time $\tau_{\alpha}$.
This is consistent with: (i) the picture that stress arises predominantly from local plasticity \cite{argon73},   
(ii) recent dielectric spectroscopy experiments indicating connections between relaxations on small and large scales \cite{hintermeyer08}, and (iii) recent NMR experiments \cite{wendlandt06} that have found the effective ``constraint'' density for deformed glasses is much larger than the entanglement density measured in the melt.
By relating strain hardening to interactions on the scale of monomers and segments, we make a novel  prediction of its magnitude.
We test our predictions using coarse-grained molecular dynamics simulations of polymer glasses, and in all cases find (at least) semiquantitative agreement.

The rest of this paper is organized as follows.  
In Section \ref{sec:theory} we motivate and develop the theory for mechanical response, make predictions that illustrate the effect of coherent relaxation in constant-strain-rate deformation and constant-strain relaxation experiments, and test these using simulations.
Finally, in Section \ref{sec:discuss}, we summarize our results, place our work in the context of recent theories and experiments, discuss how the model could be more quantitatively tested experimentally, and conclude.

\section{Theory and Simulations}
\label{sec:theory}

\subsection{Background}
\label{subsec:background}

\begin{figure}[htbp]
\includegraphics[width=3.25in]{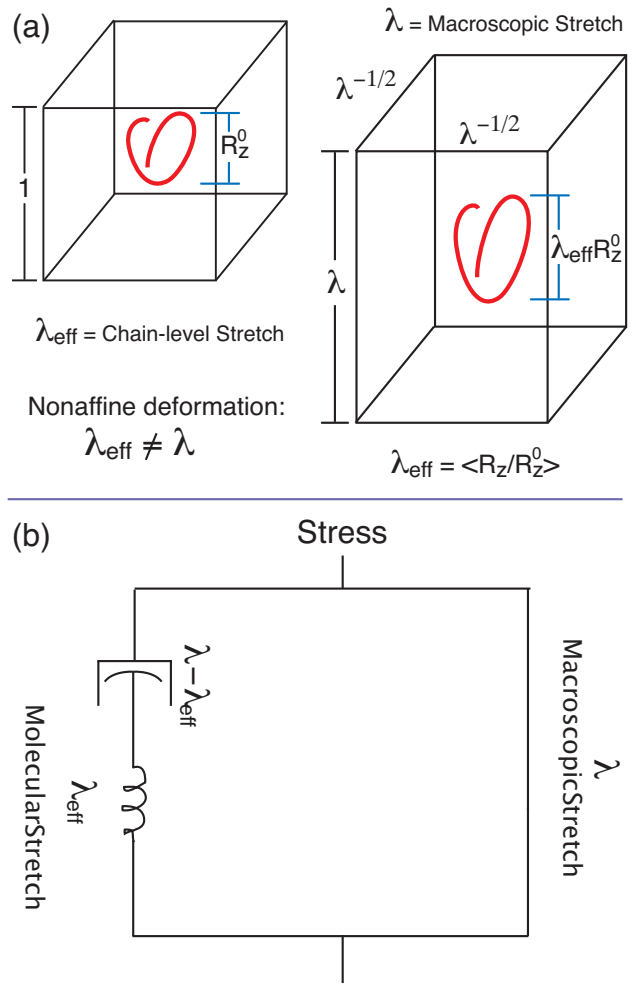}
\caption{(Color online) Schematic of our model. (a) $\lambda$ is the macroscopic stretch (observed on the scale of the experimental sample if deformation is homogeneous), while $\lambda_{eff}$ is the large-scale chain stretch.  Constant-volume `uniaxial' deformation is assumed.  (b) The spring-dashpot model for the nonaffine chain response is described by Eq.\ \ref{eq:relaxepseff}.}
\label{fig:schematic}
\end{figure}

Consider a bulk polymer sample deformed to a macroscopic stretch $\bar\lambda$.  
Classical rubber elasticity relates the decrease in entropy density to $\tilde{g}(\bar\lambda) = \frac{1}{3}(\lambda_x^2 + \lambda_y^2 + \lambda_z^2)$ \cite{treloar75, footd}.
Phenomenological `Neohookean' theories assume a strain energy density of the same form.
Both approaches give an associated (true) stress $\sigma \propto \partial \tilde{g}(\bar\lambda)/\partial\ln(\bar\lambda) \equiv g(\bar\lambda)$.
All results in this paper are presented in terms of true stresses and strains.
Stress-strain curves in well-entangled polymer glasses are often fit \cite{haward93} by 
$\sigma(\bar\lambda) = \sigma_{0} + G_R g(\bar\lambda)$, where 
$\sigma_0$ is comparable to the plastic flow stress $\sigma_{flow}$, and $G_R$ is the strain hardening modulus. 
Because of this, strain hardening has traditionally been associated \cite{arruda93b} with the change in entropy of an affinely deformed entangled network, with $G_R$ assumed to be proportional to the entanglement density $\rho_e$.
Forms for $\tilde{g}(\lambda)$ and $g(\lambda)$ for the most commonly imposed deformation modes are given in Table \ref{tab:gaussiankuhn}; Figure \ref{fig:schematic}(a) depicts the ``uniaxial'' case.

There are, however, many problems with the entropic description \cite{kramer05,hasan93,vanMelick03}.
One is that chains in uncrosslinked glasses will not 
in general deform affinely at large scales comparable to the radius of gyration.
Rather, they will possess a chain-level stretch $\bar\lambda_{eff}$   (Figure \ref{fig:schematic}(a)) which describes the deformation of chains on large scales; for example, the $zz$-component is $\left<R_z/R_z^0\right>$, where $R_z$ is the $z$-component of the rms end-to-end distance $R_{c}$, and $R_z^0$ is its value in the undeformed glass.
The deformation of well-entangled chains is consistent with an affine deformation, $\bar{\lambda}_{eff}=\bar{\lambda}$, while for unentangled systems the deformation is subaffine \cite{dettenmaier86}.
Hereon we drop tensor notation, e.\ g.\ $\bar\lambda \to \lambda$, but all quantities remain tensorial.

A key insight is that the evolution of stress in polymer glasses is controlled by the stretch $\lambda_{eff}$ of chains on scales comparable to their radius of gyration, and only indirectly by $\lambda$.  
Well below the glass transition temperature $T_g$, stress is well described \cite{hoy07,hoy09b} by 
\begin{equation}
\sigma(\bar\lambda) = \sigma_{0} + G_R^0 g(\lambda_{eff}),
\label{eq:sigmaeff}
\end{equation}
where $G_R^{0}$ is the value of $G_R$ in the long-chain limit \cite{footi}.

Equation \ref{eq:sigmaeff} shows that predicting $\lambda_{eff}$ is a critical component of the correct theory for the mechanics of uncrosslinked polymer glasses.
However, to our knowledge, no simple microscopic theory that predicts the functional form of $\lambda_{eff}$ in glasses has been published.
Most viscoelastic and viscoplastic constitutive models that describe strain hardening, e.\ g.\ Refs.\ \cite{anand03,wendlandt05}, decompose $\lambda$ into
rubber-elastic and plastic parts, or use other internal state variables, but do not explicitly account for $\lambda \neq \lambda_{eff}$ or the $N$-dependence of nonaffine relaxation \cite{footg}.
In this paper we do so; $\lambda_{eff}$ is treated as a mesoscopic (chain-level) internal state variable \cite{footg}.

\begin{table}[htbp]
\caption{Functional forms for strain hardening assuming affine constant-volume deformation by a stretch $\lambda$.}
\begin{ruledtabular}
\begin{tabular}{lccc}
Def.\ mode & $\bar{\lambda}$ & $g(\lambda)$ & $\tilde{g}(\lambda)$\\
Uniaxial & $\lambda_{x} = \lambda_{y} = \lambda_{z}^{-1/2}$ & $\lambda^{2} - \frac{1}{\lambda}$ & $\frac{1}{3}(\lambda^{2} + \frac{2}{\lambda})$\\
Plane Strain & $\lambda_{x} = \lambda_{z}^{-1},\lambda_{y}=1$ & $\lambda^{2}-\frac{1}{\lambda^{2}}$ & $\frac{1}{3}(\lambda^{2} + 1 + \frac{1}{\lambda^{2}})$
\end{tabular}
\end{ruledtabular}
\label{tab:gaussiankuhn}
\end{table}

\subsection{Maxwell-like model for $\lambda_{eff}$}
\label{subsec:maxwell}

We now develop a predictive theory for $\lambda_{eff}$.  
The problem is most naturally formulated in terms of true strains $\epsilon_{eff} = \ln(\lambda_{eff})$ and $\epsilon = \ln(\lambda)$.
We postulate a simple Maxwell-like model for the relaxation of $\epsilon_{eff}$.
The governing equation for the model shown in Figure \ref{fig:schematic}(b) is
\begin{equation}
\dot{\epsilon}_{eff} = \dot{\epsilon} - \epsilon_{eff}/\tau.
\label{eq:relaxepseff}
\end{equation}
Equation \ref{eq:relaxepseff} is a standard ``fading memory'' form implying chains ``forget'' their large-scale orientation at a rate $\tau^{-1}$.
In other words, $\tau$ is the time scale over which $\epsilon_{eff}$ will relax towards its ``equilibrium'' value $\epsilon_{eff} = 0$ (we assume that chains are not crosslinked).

In this formulation $\epsilon_{eff}$ corresponds to the strain in the ``spring'' in Fig.\ \ref{fig:schematic}(b).
However, the stress we will associate with increasing $|\epsilon_{eff}|$ is \textit{viscoelastoplastic}. 
Here $\epsilon_{eff}$ is an (in principle) \textit{micro}-reversible strain corresponding to chain orientation; it is not an elastic strain in the macroscopic sense of shape recoverability of a bulk sample \cite{foot5}.
$\epsilon_{pl} \equiv \epsilon - \epsilon_{eff}$ corresponds to the ``dashpot'' strain used in many constitutive models; it is plastic in the sense of being both micro- and macro-irreversible.

Maxwell-like models have been used to describe polymer viscoelasticity for more than half a century, and complicated ladder models were developed (e.\ g.\ Ref.\ \cite{gross56}) because of the inadequacy of earlier single rate models.
However, we will provide evidence below that the correct choice of meso-variable (i.\ e.\ $\lambda_{eff}$ or $\epsilon_{eff}$) restores the applicability of a single (albeit $N$-dependent) relaxation time model, at least for monodisperse systems.

\subsection{Coherent strain-activated relaxation}
\label{subsec:coher}

The next step in constructing a useful microscopic theory is prediction of $\tau$.
If one supposes that chains under active deformation at strain rate $\dot\epsilon$ relax \textit{coherently}, and that the relaxation is strain-activated, $\tau$ is reduced by a factor of $N\dot\epsilon$ relative to its quiescent value.

We assume relaxation on large scales is coupled to segmental relaxation.
For an \textit{a prirori} unspecified relaxation dynamics in the quiescent state, 
\begin{equation}
\begin{array}{cc}
\tau \sim N^{\gamma} \tau_{\alpha} & \textrm{(incoherent)},
\end{array}
\label{eq:tauincoh}
\end{equation} 
where $\tau_{\alpha}$ is the  ``alpha'' or segmental relaxation time, and $\gamma$ is unspecified and may be $N$-dependent.

Gaussian polymers have chain statistics defined by $R_c^{2} \equiv R_{s}^{2} N$, where $R_s^2$ is the squared statistical segment length.
Mathematically, this gives the identity
\begin{equation}
\partial R_c^{2}/\partial t = N\partial R_s^{2}/\partial t.
\label{eq:factorofN}
\end{equation}
While Eq.\ \ref{eq:factorofN} surely oversimplifies the physics of glasses (e. g. it does not hold in quiescent systems because $R_c^{2}$ is stationary), nonetheless it suggests an associated relaxation rate under active deformation that is $N$ times larger (or a time $N$ times smaller) than the value in the quiescent state:
\begin{equation}
\begin{array}{cc}
\tau \sim N^{\gamma-1} \tau_{\alpha} & \textrm{(coherent)}.
\end{array}
\label{eq:taucoh}
\end{equation} 
We postulate that Eq.\ \ref{eq:factorofN} becomes valid in actively deformed glasses; segmental rearrangements become ``coherent'' because rearrangements  that restore $R_c^2$ towards its initial value dominate over those which do not.
In practice, coherent relaxation is forced by the stiffness of the covalent backbone bonds, which have (nearly) constant length $l_0$ and so maintain (nearly) constant chain contour length $L = (N-1)l_0$.

Recent work assists in hypothesizing a more specific relaxation dynamics for actively deformed systems.
Ref.\ \cite{hoy09b} showed that chains in not-too-densely \cite{footq} entangled model polymer glasses well below $T_g$ orient independently of one another during active deformation.
The behavior observed was that of individual chains coupled to a ``mean-field'' glassy medium.
If relaxations on chain and segmental scales are \textit{tightly} coupled, and chains relax independently of one another, $\gamma = 2$ is predicted \cite{doi86}.
Then
\begin{equation}
\begin{array}{cc}
\tau \sim N^{2} \tau_{\alpha} & \textrm{(incoherent)},\\
& \\
\tau \sim N \tau_{\alpha} & \textrm{(coherent)}.
\end{array}
\label{eq:taucohincoh}
\end{equation}

Within the framework of Eqs.\ \ref{eq:tauincoh}, \ref{eq:taucoh}, and \ref{eq:taucohincoh}, when relaxation is strain activated, both $\tau_{\alpha}$ \cite{chen07} and $\tau$ will be reduced by a factor $\sim \dot\epsilon$, so the overall reduction in $\tau$ during active deformation scales as $N\dot\epsilon$.  
Note that the scaling analysis presented above does not account for additional changes in $\tau_{\alpha}$ arising from other causes, e.\ g.\ increased mobility associated with yield.
Also note that the above arguments assume $\epsilon_{eff}$ and $\dot\epsilon$ have the same sign.
If deformation is reversed, e.\ g.\ in a Bauschinger-effect experiment \cite{senden10,ge10}, and the sign of $\dot\epsilon$ is opposite that of $\epsilon_{eff}$, active deformation may not produce coherent relaxation.

\subsection{Bead-spring simulations}

Some of the arguments made in Sections \ref{subsec:maxwell}-\ref{subsec:coher} were heuristic, and several assumptions were made, so it is important to compare the theoretical predictions with results from simulations.
For example, observations of sharp changes in segmental relaxation times with strain and significant dynamical heterogeneity \cite{lee09,lee09b,riggleman08} during deformation are seemingly at odds with our postulated single, constant relaxation time $\tau$.
Further, entangled chains may not be able to relax coherently if the entanglements concentrate stress, so the reduction of $\tau$ in actively deformed entangled systems may be weaker than predicted above.

The basic ideas presented above can be tested using molecular dynamics simulations of the Kremer-Grest bead spring model \cite{kremer90}. 
Polymer chains are formed from $N$ monomers of mass $m$.  
All monomers interact via the truncated and shifted Lennard-Jones potential $U_{LJ} = 4u_0\left[(a/r)^{12} - (a/r)^{6} - \left((a/r_c)^{12} - (a/r_c)^{6}\right)\right]$.
Here $r_c = 1.5a$.
Covalently bonded monomers additionally interact via the FENE potential $U_{FENE} = -(kR_0^{2}/2)\ln(1 - (r/R_0)^2)$; the canonical \cite{kremer90} values $k = 30u_0/a^2$ and $R_0 = 1.5a$ are employed.
All quantities are expressed in terms of the intermonomer binding energy $u_0$, monomer diameter $a$, and characteristic time $\tau_{LJ} = \sqrt{ma^2/u_0}$.    
The equilibrium covalent bond length is $l_0 = .96a$ and the Kuhn length in the melt state is $l_K = 1.8a$.

All systems have $N_{ch}$ chains, with $N_{ch}N \simeq 2.5\times10^{5}$. 
Periodic boundary conditions are applied along all three directions of the simulation cell, which has periods $L_x,\ L_y,\ L_z$ along the $x,\ y,\ z$ directions.
Melts are equilibrated \cite{auhl03} and rapidly quenched ($k_B\dot{T} = -.002u_0/\tau_{LJ}$) into glasses at $T=0.2u_0/k_B \simeq 0.6T_g$.
Uniaxial-stress compressive deformations are then imposed, using the same protocols employed in previous work \cite{hoy07}; a constant true strain rate $\dot\epsilon = \dot{L}_z/L_z = -10^{-5}/\tau_{LJ}$ is applied, with $\lambda = L_z/L_z^{0}$.
A Langevin thermostat with damping time $10\tau_{LJ}$ is used to maintain $T$, and a Nose-Hoover barostat with damping time $100\tau_{LJ}$ is used to maintain zero pressure along the transverse directions.
The values of $|\dot{\epsilon}|$ and $T$ employed here
lie within ranges shown \cite{rottler03c,robbins09} to reproduce many experimental trends \cite{haward97}, such as logarithmic dependence of $\sigma$ on $\dot\epsilon$ and linear scaling of the hardening modulus with the flow stress.

\subsection{Chain conformations under deformation and constant-strain relaxation}

Constant strain-rate deformation and constant strain relaxation are two of the most commonly performed mechanical experiments.
In a constant strain rate experiment, assuming $\tau$ is independent of $\epsilon$, i.\ e.\ assuming polymer glasses are \textit{linearly} viscoplastic, the solution to Eq.\ \ref{eq:relaxepseff} is \cite{foot1}
\begin{equation}
\epsilon_{eff}(\epsilon) =  \dot{\epsilon}\tau\left(1-\exp(-\epsilon/\dot\epsilon\tau)\right) \equiv \dot{\epsilon}\tau\left(1-\exp(-t/\tau)\right).
\label{eq:epseffoft}
\end{equation}
If a system is deformed to a strain $\epsilon^{0}$ and effective strain $\epsilon_{eff}^{0}$, and then deformation is ceased, Eq.\ \ref{eq:relaxepseff} has the solution 
\begin{equation}
\epsilon_{eff} = \epsilon_{eff}^{0}\exp(-t/\tau).
\label{eq:relax}
\end{equation}
In this case, our model predicts slowdown in relaxation upon cessation of deformation; the $\tau$ in Eq.\ \ref{eq:relax} is $N\dot\epsilon^{-1}$ times larger than the $\tau$ in Eq.\ \ref{eq:epseffoft}.

\begin{figure}[htbp]
\includegraphics[width=3.25in]{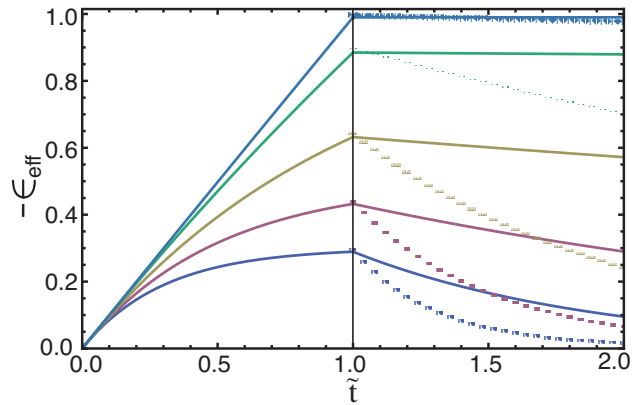}
\caption{(Color online) Compressive deformation followed by relaxation at constant strain.  
Curves from top to bottom are for $N=500$, $N=40$, $N=10$, $N=5$, and $N=3$.  
$\tilde{t} = |\dot\epsilon| t$ is time scaled by the strain rate applied for $0 \leq \tilde{t} \leq 1$.    
For $\tilde{t} < 1$ curves show predictions of Eq.\ \ref{eq:epseffoft}, while for $\tilde{t} > 1$ curves and symbols show predictions of Eq.\ \ref{eq:relax}.
Solid curves assume $\tau \propto N^{\gamma-1}$ during deformation and $N^{\gamma}$ at constant strain, while symbols assume $\tau \propto N^{\gamma-1}$ at all $\tilde{t}$.  Here $\gamma = 2$.}
\label{fig:goodbadrelax}
\end{figure}

Figure \ref{fig:goodbadrelax} shows theoretical predictions of Eqs.\ \ref{eq:epseffoft}-\ref{eq:relax} for evolution of $\epsilon_{eff}(t)$ in systems compressively strained to $\epsilon = -1.0$ at constant rate (for $0 < t < \dot\epsilon^{-1}$), and then allowed to relax at constant strain.
Results are plotted against $\tilde{t} = \dot\epsilon t$, where $\dot\epsilon$ is the strain rate applied during compression.
The solid lines assume $\gamma = 2$, and relaxation is coherent ($\tau \propto \dot\epsilon^{-1}N^{\gamma-1}$) during active deformation and becomes incoherent ($\tau \propto N^{\gamma}$) when deformation is ceased, consistent with our theoretical picture.
For the purpose of contrast, the symbols show predictions assuming that relaxation remains coherent (i.\ e.\ $\tau$ increases by only a factor $\dot\epsilon^{-1}$) after deformation is ceased.
$N = 500$ chains orient nearly affinely during strain: $\epsilon_{eff} \simeq \dot\epsilon t$, while short chains orient much less.
The solid lines are far more consistent with both experiments \cite{loo00,lee09,lee09b} and simulations \cite{capaldi02,riggleman08}, which show relaxation slows dramatically upon cessation of active deformation.

\begin{figure}[htbp]
\includegraphics[width=3.25in]{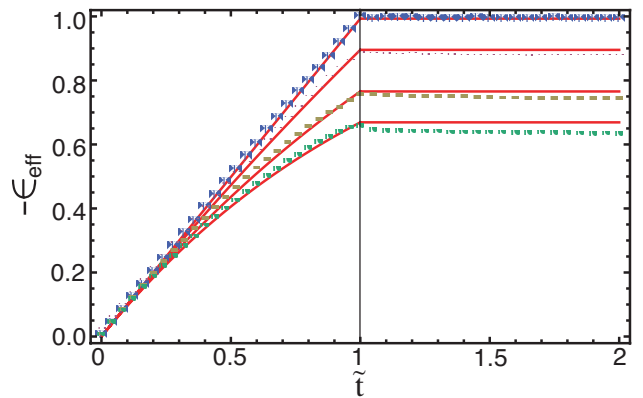}
\caption{(Color online) $\epsilon_{eff}$ vs.\ $\tilde{t}$ for uniaxial compression followed by constant strain relaxation; comparison of theory and simulation results.   Symbols from bottom to top are bead-spring simulation results for $N = 500$, $36$, $18$, and $12$.  For $\tilde{t} \leq 1$, lines are fits to Eq.\ \ref{eq:epseffoft}; the fit values of $\tau$ are given in Table \ref{tab:tauovN}.    For $\tilde{t} \geq 1$ the lines are given by Eq.\ \ref{eq:relax}; no further fitting is employed, and the values of $\tau$ from Table \ref{tab:tauovN} are multiplied by $N\dot\epsilon^{-1}$, consistent with the transition from coherent to incoherent relaxation assumed to occur upon cessation of deformation.
}
\label{fig:lameffoft}
\end{figure}

Figure \ref{fig:lameffoft} shows simulation data for $\epsilon_{eff}(t)$ under the same procedure of compression followed by constant-strain relaxation.
Solid lines for $\tilde{t} \leq 1$ are fits to Eq.\ \ref{eq:epseffoft}.
Values for $\tau$ from these fits are given in Table \ref{tab:tauovN}.  
The data are quantitatively consistent with $\tau \propto N^{\gamma-1}$ and $\gamma = 2$.
In particular, values for $\tau/(N-1)$ are nearly constant.
For this model the entanglement length is $N_e \simeq 85$ \cite{hoy09}, so this trend spans the range from  unentangled to well-entangled chains.
The effect of entanglements is more consistent with an increase in the prefactor of $\tau/(N-1)$ than a change in $\gamma$.

As shown in Table \ref{tab:tauovN}, the agreement of $\tau$ with 
the prediction $\tau_\alpha \simeq .1\dot{\epsilon}^{-1}$ under active deformation \cite{chen07}, where we identify $\tau_\alpha = \tau/(N-1)^{\gamma-1}$, is quantitative.
The data in Fig.\ \ref{fig:lameffoft} is consistent with our hypothesis that $\tau$ remains constant during deformation; only small deviations from the fits to Eq.\ \ref{eq:epseffoft} are apparent on the scale shown.
Given the large variation in stress as strain increases (see Section \ref{subsec:stressstrain}), it is remarkable that a single relaxation rate theory fits $\epsilon_{eff}$ so well.
However, this is consistent with the enhancement of dynamical homogeneity and narrowing of the relaxation spectrum under flow observed in similar models in Refs.\ \cite{riggleman08,warren10}. 
In developing our theory for the mechanical response, we will assume for simplicity that $\tau$ is indeed independent of $\epsilon$.

\begin{table}[htbp]
\caption{Values of $\tau$ obtained by fitting $\epsilon_{eff}(t)$ to Eq.\ \ref{eq:epseffoft} (with $\gamma = 2$) as a function of chain length $N$, for flexible bead spring chains.  Statistical uncertainties on the fits, i.\ e.\ the error bars for $\tau$, increase with $N$, to $\sim 5\%$ for $N=500$.  The factor of $N-1$ in the rightmost column arises because in a discrete-bead model, relaxation is associated with the number of covalent bonds rather than monomers.  The data is from Fig.\ \ref{fig:lameffoft}.  Note $\dot\epsilon = 10^{-5}/\tau_{LJ}$ so $\dot\epsilon\tau$ is of order unity for chains with $N \sim 10$.  For this model $N_e \simeq 85$.}
\begin{ruledtabular}
\begin{tabular}{ccc}
$N$ & $\tau/\tau_{LJ}$ & $\tau/((N-1)\tau_{LJ})$ \\
12 & $1.15\cdot10^{5}$ & $1.05\cdot10^{4}$ \\
18 & $1.78\cdot10^{5}$ & $1.05\cdot10^{4}$ \\
36 & $4.49\cdot10^{5}$ & $1.09\cdot10^{4}$\\
71 & $7.19\cdot10^{5}$ & $1.03\cdot10^{4}$ \\
107 & $1.12\cdot10^{6}$ & $1.05\cdot10^{4}$ \\
250 & $2.76\cdot10^{6}$ & $1.11\cdot10^{4}$ \\
500 & $7.14\cdot10^{6}$ & $1.43\cdot10^{4}$\\
\end{tabular}
\end{ruledtabular}
\label{tab:tauovN}
\end{table}

After cessation of deformation, simulation results in Fig.\ \ref{fig:lameffoft} suggest Eqs.\ \ref{eq:relaxepseff} and \ref{eq:relax} give a qualitatively valid description of large scale chain relaxation at constant strain.
The small initial decreases in $|\epsilon_{eff}|$ at times ($\tilde{t} = 1+\delta,\ \delta \ll 1$) are larger than the theoretical predictions of Eq.\ \ref{eq:relax} for small $N$. 
This may be attributable to chain end or segmental-relaxation effects not included in our model.
At larger $\tilde{t}$, however, it is clear that the increase in $\tau$ upon cessation of deformation is qualitatively captured.

\subsection{Microscopic theory for viscoplastic stress}
\label{subsec:stresstheory}

Here we will derive a theory of Neohookean viscoplasticity for the $N$-dependent stress-strain response.  
The work to deform the system is broken into two components corresponding to an isotropic resistance to flow and an anisotropic resistance to chain orientation.
This general approach has been employed many times before, e.\ g.\ in Refs.\ \cite{boyce88, arruda93b}. 
We present a novel microscopic picture that quantitatively associates Gaussian/Neohookean strain hardening with the energy dissipated in local segmental hops, and quantitatively associates the smaller hardening observed for shorter chains to strain-activated relaxation of large-scale chain orientation (i.\ e.\ $\lambda \neq \lambda_{eff}$).

From the second law of thermodynamics, the work $W$ required to deform a polymer glass is $W(\epsilon) = \Delta E(\epsilon) + \Delta Q(\epsilon)$, where $E$ is the internal energy and $Q$ is the portion of work converted into heat, including both dissipative and entropic terms.
Without loss of generality, this can be rewritten as $W(\epsilon) = W_1(\epsilon) + W_2(\epsilon)$, where 
$W_2$ is the viscoplastic component of the work.
$W_1$ captures ``everything else'', such as elastic terms pre-yield, and (in principle, though they are not treated in this paper) other effects such as softening, anelastic energetic stresses, and entropic stresses. 
In experiments, ductile deformation of glassy polymers occurs at nearly constant volume \cite{haward97}; thus we assume constant volume deformation and  
treat $W_1$ and $W_2$ as intensive quantities.

Both rubber elasticity and Neohookean elasticity associate strain hardening with $W_1$.
However, experiments and simulations \cite{hasan93,hoy07} have shown that $W_2$ is the dominant term for strains ranging from the beginning of the plastic flow regime to the onset of dramatic ``Langevin'' \cite{arruda93b} hardening.
The latter occurs at very high strains for most synthetic polymers \cite{haward93} and has been associated with the increase in energy arising from chain stretching between entanglements \cite{chui99,hoy07}.
Simulations have provided strong evidence that in this regime, $W_2$ is closely connected with the same local interchain plastic events that control the flow stress \cite{hoy07}.
These events have a characteristic energy density $\sim u_0/a^{3}$, where $u_0$ is the energy scale of secondary (i.\ e.\ noncovalent) interactions \cite{footy}. 
Here we treat the regime where $W_2$ dominates, and neglect $W_1$, i.\ e.\ we make the approximation $W = W_2$.

$W$ may be further broken down into  ``segmental'' and ``polymeric'' terms: $W = W^{s} + W^{p}$, where $W^{s}$ accounts for the plastic flow stress in the absence of hardening, and $W^{p}$ accounts for the viscoplastic component of strain hardening.
Since this paper focuses on strains well beyond yield, for convenience we choose a standard viscous-yield term
\begin{equation}
W^{s} = \displaystyle\frac{u_0}{a^{3}}\left(\epsilon + \epsilon_y \rm{exp}(-\epsilon/\epsilon_y)\right),
\label{eq:w2s}
\end{equation}
where $\epsilon_y$ is the yield strain.

The stress is given by $\sigma = \partial W/\partial\epsilon$.
It can similarly be written as a sum of segmental and `polymeric' contributions 
\begin{equation}
\sigma = \sigma^{s} + \sigma^{p} = \partial W^{s}/\partial\epsilon +  \partial W^{p}/\partial\epsilon.
\label{eq:sig2parts}
\end{equation}
Ref.\ \cite{hoy07} showed $\sigma \propto R_p$, where $R_p$ is the rate (per unit strain) of plastic events identified by local rearrangements.
The natural correlation length scale for the local plastic rearrangements, 
since according to the above arguments they are coherent, is $R_c$, and the associated volume is $V = R_c^{3}/6^{3/2}$. 
Here factors of $\sqrt{6}$ arise from the standard relation for the radius of gyration $R_g$ of Gaussian polymers, $R_g = R_c/\sqrt{6}$.

We will now associate polymeric strain hardening with $W^{p}$ and replace $\lambda$ by $\lambda_{eff}$ in order to calculate an $N$-dependent $\sigma^{p}$.
We postulate that $W^{p}$ is controlled by the increase in $V$; in other words, strain hardening occurs because the volume $V$ controlling $W^{p}$ increases faster than it can relax \cite{footl}.
Studies of bidisperse mixtures have provided strong evidence that the evolution of $\lambda_{eff}$ (and hence strain hardening) can be understood in terms of single chains interacting with a glassy mean field \cite{hoy09b}, so it is natural to assume the plastic events are ``unary'' (in the sense that $\geq2$-chain effects are unimportant). 
$W^{p}$ will then scale linearly with $\rho_{cr}$, where 
\begin{equation}
\rho_{cr} = \sqrt{6}\rho l_0/(NR_{c})
\label{eq:rhocrdef}
\end{equation} 
is the density of coherently relaxing contours, and $\rho$ is monomer number density.
Then, incrementally
\begin{equation} 
\Delta W^{p} \simeq (u_0/a^{3})\Delta (\rho_{cr} V) = (u_0/a^{3}) N^{-1} \rho l_0 \Delta(R_{c}^2)/6.
\label{eq:deltawp}
\end{equation} 
Recall $R_{c}^2 = 3l_0 l_K N \tilde{g}(\lambda_{eff})$, where $l_K$ is (here) the ``equilibrium'' Kuhn length in the undeformed glass.  Then for a deformation increment $\Delta\bar{\lambda}$,
\begin{equation}
\Delta(R_{c}^2) = 3l_0 l_KN\ \Delta \left(\tilde{g}(\lambda)\big{|}_{\lambda_{eff}}\right),
\label{eq:deltarc2}
\end{equation}
The term in parentheses indicates the difference is evaluated at $\lambda = \lambda_{eff}$.

Combining equations \ref{eq:deltawp}-\ref{eq:deltarc2} gives 
\begin{equation}
\Delta W^{p} =  (u_0/a^{3}) \rho l_0^{2} l_K \Delta(\tilde{g}(\lambda_{eff}))/2.
\label{eq:deltawp2}
\end{equation}
This result combined with Eq.\ \ref{eq:w2s} gives a prediction for the stress:
\begin{equation}
\sigma(\epsilon_{eff}) = \displaystyle\frac{u_0}{a^3}\left(1 - \exp(-\epsilon/\epsilon_y) + \rho l_0^{2} l_K |g(\epsilon_{eff})|/2\right),
\label{eq:hardeningstress}
\end{equation}
where 
$g(\epsilon) = (3/2) \partial \tilde{g}/\partial \epsilon$.  
The absolute value $|g(\epsilon_{eff})|$ appears because we have so far treated $\sigma$ as positive.

Eq.\ \ref{eq:hardeningstress} has several interesting features, which we now relate to previous models.
First, it predicts that $\sigma^{p}$ (at fixed strain) increases with increasing $l_K$.
This is consistent with the well-established result that straighter chains are harder to plastically deform \cite{haward93,haward97}. 
However, the power of $l_K$ on which $\sigma^{p}$ depends is sensitive to our theoretical assumptions, specifically the definition of $\rho_{cr}$ (Eq.\ \ref{eq:rhocrdef}).
Other definitions can produce $G_R \propto l_K^{3/2}$ or $l_K^{3}$, but choosing which one is ``best'' \cite{footn} requires greater knowledge of the variation in local plasticity (at a microscopic level) with $l_K/l_0$ than is currently available. 
We test Eq.\ \ref{eq:hardeningstress} using MD simulations in the following section.
While only one value of $l_K/a$ is considered here, it is large enough (1.8) that viscoplastic contributions to $\sigma$ scaling as $l_K^{3}$ would be much larger than contributions scaling as $l_K$.

Second, for long chains, our theory predicts little relaxation during deformation, and a strain hardening modulus $G_R \simeq \rho l_0^{2}l_K$.  
This value is much closer to the effective constraint density measured in (NMR) experiments of deformed glassy samples \cite{wendlandt06} than to the entropic prediction $G_R = \rho_e k_BT$.
It must be noted that increasing $l_K$ also increases $\rho_{e}$ \cite{fetters99b}.
However, considerable evidence (e.\ g.\ \cite{hasan93,senden10,ge10,chen09,hine07,hoy09b,footw})
suggests that chain orientation and local, secondary interactions which act over scales $\sim (l_0 \sim a \sim l_K)$ are the true controlling factors for $\sigma$, at least during the initial stages of hardening.
Finally, Haward postulated that $G_{R}$ arises from the contraints imposed by the mesh of uncrossable chains \cite{haward93}; our argument that hardening scales with $\rho_{cr}$ are consistent with this hypothesis.

\subsection{Predictions for stress-strain curves}
\label{subsec:stressstrain}

In Equation \ref{eq:hardeningstress} we have assumed that all contributions to $\sigma$ scale with a single energy density (i.\ e.\ stress) $(u_0/a^{3})$. 
This is an approximation, since flow and hardening stresses have been observed to be linearly rather than directly proportional \cite{govaert08}.
However, the `constant offset' term in this linear relationship is often fairly small compared to the linear term (especially for $T$ well below $T_g$ \cite{govaert08,hoy09b}), so the approximation is reasonable.
Therefore we associate $u_0/a^{3}$ with $\sigma_{flow}$, and scale it out.  This gives
\begin{equation}
\sigma^{*}(\epsilon_{eff}) = A\displaystyle\frac{\sigma(\epsilon_{eff})}{u_0/a^{3}},
\label{eq:stresseps}
\end{equation}
where $A$ is a prefactor arising from our neglect of prefactors in the above analysis.
Note that all ``thermal'' aspects of our theory are implicitly wrapped into $A$ and $\tau_{\alpha}$. 
In both real and simulated systems $\sigma_{flow}$ and $A$ are approximately proportional to ($1 - T/T_g$) \cite{argon73, rottler03c, vanMelick03,hoy06}, so this scaling should remove much of the $T$-dependence.
The merits of ``multiplicative'' forms like Eq.\ \ref{eq:stresseps} for predicting stress have been discussed recently in Refs.\ \cite{senden10, wendlandt10}.
More sophisticated models (e.\ g.\ Ref.\ \cite{chen09}) explicitly treat the variation of $A$ with $\dot\epsilon$ and $T$, and/or the variation of $\tau_{\alpha}$ with $\sigma$, $T$, and local microstructure.
Here, however, $A$ and (for fixed $N$) $\tau_{\alpha}$ are treated as numerical constants. 

\begin{figure}[htbp]
\includegraphics[width=3.25in]{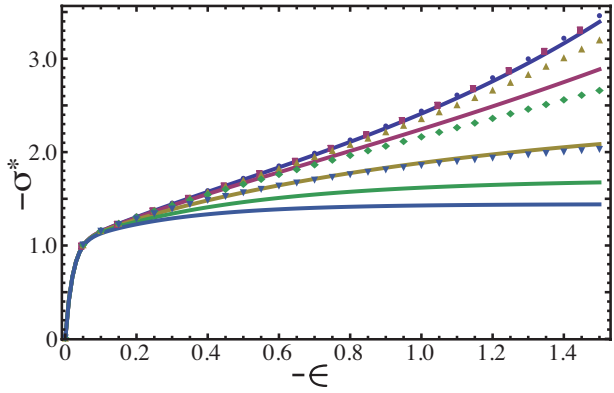}
\caption{(Color online) Stress-strain curves predicted by Eqs.\ \ref{eq:stresseps}-\ref{eq:g1} with $A=1$.   $-\sigma^{*}$ and $-\epsilon$ are shown because stress and strain are negative for compression. Curves from top to bottom are for $N=500$,  $N=10$, $N=5$, and $N=3$.  Predictions in the $N\to \infty$ limit are not distinguishable from $N=500$ predictions on the scale of this plot.  Solid curves assume $\dot\epsilon\tau = B N^{\gamma-1}$, while symbols assume $\dot\epsilon\tau = B  N^{\gamma}$.  In both cases, $\gamma = 2$, and as suggested by Ref.\ \cite{chen07}, $B=0.1$.}
\label{fig:conststrainrate}
\end{figure}

To predict the $N$-dependence of $\sigma(\epsilon)$, an analytic form for $g(\epsilon_{eff}(\epsilon))$ is obtained by plugging in the solution for $\epsilon_{eff}(\epsilon) $ from Eq.\ \ref{eq:epseffoft} into the form of $g(\epsilon)$ for uniaxial deformation (Table \ref{tab:gaussiankuhn}, with $\exp(\epsilon) = \lambda$):
\begin{equation}
\begin{array}{lcc}
g(\epsilon_{eff}(\epsilon)) & = & \textrm{exp}\left[2 \dot\epsilon \tau  (1-\textrm{exp}(-\epsilon/\dot\epsilon \tau))\right]\\
 &- & \textrm{exp}\left[-\dot\epsilon \tau (1- \textrm{exp}(-\epsilon/\dot\epsilon \tau))\right].
\end{array}
\label{eq:g1}
\end{equation}   
Figure \ref{fig:conststrainrate} shows predictions of Eqs.\ \ref{eq:stresseps}-\ref{eq:g1} for 
$\sigma^{*}(\epsilon)$ in uniaxial compression at various $N$, with $A=1$, $|\epsilon_y = .02|$, $\rho = 1.0a^{-3}$, $l_0 = .96a$, and $l_K = 1.8a$; the latter three are chosen to match the flexible bead spring model employed in the simulations.
Solid lines assume $\tau\propto N^{\gamma-1}$ with $\gamma = 2$ as discussed above, while symbols assume coherent chain relaxation is not important and $\tau\propto N^{\gamma}$.
The solid lines are qualitatively consistent with simulations \cite{hoy06,hoy07},
while the symbols are inconsistent.  
Both show increasing strain hardening with increasing $N$, but in the latter case hardening increases much faster and saturates at a much lower value of $N$ than is realistic. 
For example, the $\tau\propto N^{\gamma}$ predictions for $N = 40$ and $N = 500$ are indistinguishable on the scale of the plot.
Results for tension are not presented here because our model includes no asymmetry between tension and compression \cite{footp}.

The large-strain $(\epsilon \gg \epsilon_y$) mechanical response predicted by our model varies continuously from perfect-plastic flow ($\sigma^{*}(\epsilon) \to$ a constant $\sigma_{flow}$) to network-like polymeric response ($\sigma^{*}(\epsilon) \to 1 + \rho l_0^{2}l_K g(\epsilon)$) as $\dot\epsilon\tau$ varies from zero to  $\infty$ (equivalently, as $N$ increases).
In betweeen these limits, the response is ``polymeric viscoplasticity''.
Note that since we treat $\epsilon_{eff}$ as microreversible, a more accurate term for our model is ``viscoelastoplasticity'' \cite{foot3}.
However, since our model does not treat stress relaxation after cessation of deformation, $\sigma^{*}$ may be regarded as a (reduced) orientation-dependent plastic flow stress.  

\begin{figure}
\includegraphics[width=3.25in]{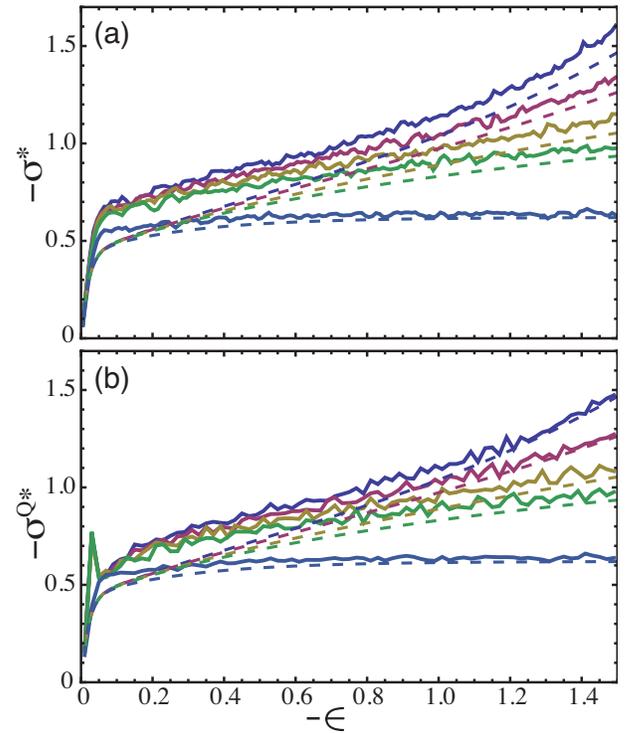}
\caption{(Color online) Stress-strain curves ($\sigma^{*}$) from bead spring simulations (solid lines) and theoretical predictions of Eqs.\ \ref{eq:stresseps}-\ref{eq:g1}  [(a) $\sigma^{*}$, (b) $\sigma^{Q*}$].  
Values of $N$ from top to bottom are $500$, $36$, $18$, $12$, and $4$.  The top 4 are from the same simulations shown in Fig.\ \ref{fig:lameffoft} and use the fit values of $\tau$ (Table \ref{tab:tauovN}), while the bottom theory curve assumes $\dot\epsilon\tau = .1(N-1)$.}
\label{fig:stresses}
\end{figure}

We now compare theoretical predictions for stress-strain curves to results from bead-spring simulations.
In Figure \ref{fig:stresses}, dashed lines show predictions of Eqs.\ \ref{eq:stresseps}-\ref{eq:g1} with $A = 0.43$, values of $\tau$ taken from Table \ref{tab:tauovN}, and the same values of $|\epsilon_y|$, $l_0$ and $l_K$ as in Fig.\ \ref{fig:conststrainrate}.
The value of $A$ obtained in Fig.\ \ref{fig:stresses} is comparable to $(1-T/T_g)$.
Stress-strain curves from simulations are shown as solid lines.
Note that stresses and strains are negative in compression.
Panel (a) shows $\sigma^{*}$, while panel (b) shows its dissipative component $\sigma^{Q*} = (\sigma^{*} -  (u_0/a^{3})^{-1}\partial U/\partial\epsilon)$ \cite{hoy07}.  
While this definition of $\sigma^{Q*}$ includes entropic terms, these are known to be only of order $1\%$ of the total stress at this $T$ \cite{hoy07}.

In both panels, the correct trends are predicted, and quantitative agreement is within $\sim20\%$.
Very short chains ($N=4$) show nearly perfect-plastic flow, while longer chains show strain hardening, similar to results analyzed in many previous studies.
The quantitative differences at small strains arise primarly from our oversimplified treatment of yield \cite{foot2}. 
At large strains, comparing panels (a-b), it is apparent that differences between predictions of Eqs.\ \ref{eq:stresseps}-\ref{eq:g1} and simulation results arise largely from energetic terms associated with strain hardening, i.\ e.\ covalent bond energy and additional plastic deformation arising from chain stretching between entanglements.
Experiments on amorphous polymer glasses (e.\ g.\ \cite{hasan93}) have shown a similar percentage of the total stress 
is associated with energetic terms, strain softening, etc., so the agreement between theoretical predictions and bead-spring results is satisfactory given the simplicity of our model.
In panel (b), the ``spike'' in simulation results at small strains for $N>4$ reflects yield and subsequent strain softening.

\subsection{Nonaffine displacement and plastic deformation}
\label{subsec:NAD}

\begin{figure}[htbp]
\includegraphics[width=3.25in]{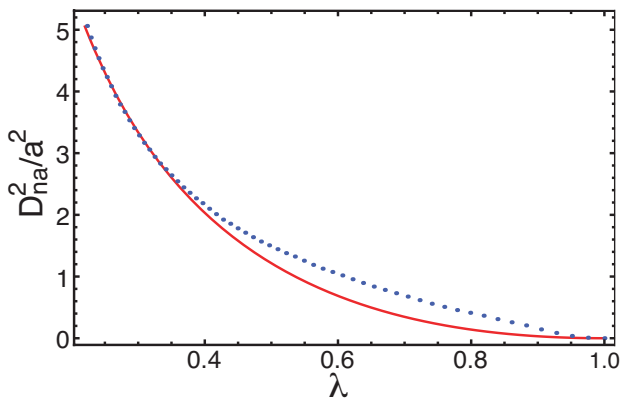}
\caption{(Color online) Nonaffine displacement at low $T$ in systems of long chains.  Circles show data from bead-spring simulations (with $N=500$ and $T=.01u_0/k_B$).  The solid line shows a fit to $D^{2}_{na}/a^{2} = C(S(\lambda)-1)$ (see Eq.\ \ref{eq:soflam}) with $C=7.39$.  Note that the maximum value of $D^{2}_{na}$ is well below the squared tube diameter $d_T^{2}/a^{2} \sim 100$ \cite{doi86,everaers04}, consistent with the picture that entanglements do not control the mechanical response of these systems.}
\label{fig:NAD}
\end{figure}

We have argued above that coherent relaxation is driven by the stiff covalent bonds and the need to maintain constant chain contour length $L$.
Since long chains have $\dot\epsilon\tau \gg 1$ and deform affinely on the end-to-end scale, they must deform nonaffinely on smaller scales to maintain chain connectivity.
It is interesting to relate this nonaffine deformation to the chain stretching that would occur if deformation were affine on all scales and $L$ was not constrained, i.\ e.\ the limit of a Gaussian coil with zero spring constant \cite{doi86} embedded in a deforming medium.
For uniaxial tension or compression the nonaffine displacement should be given by $D^{2}_{na} \sim S(\lambda) -1$, with
\begin{equation}
S(\lambda) \equiv  \displaystyle\frac{L(\lambda)}{L(1)} = \displaystyle\frac{1}{2}\left(\lambda + \displaystyle\frac{\sin^{-1}(\sqrt{1 - \lambda^3})}{\sqrt{\lambda - \lambda^{4}}}\right).
\label{eq:soflam}
\end{equation}
Figure \ref{fig:NAD} shows data for the squared nonaffine displacement of monomers, $D^{2}_{na} = \left<(\vec{r} - \bar{\lambda}\vec{r}_{0})^2\right>$, where the monomer positions are $\{\vec{r}\}$ at stretch $\bar\lambda$ and $\{\vec{r}_0\}$ in the initial state, for $N = 500$.
Data for $T = .01u_0/k_B$ is shown to minimize the thermal contribution to $D^{2}_{na}$.
A fit to $S(\lambda)-1$ is also displayed. 
There is qualitative agreement at large strains $(\lambda \ll 1)$, and the underestimation of $D^{2}_{na}$ at smaller strains is largely attributable to smaller scale (i.\ e.\ incoherent) plasticity on scales approaching the monomer diameter $a$.
Another reasonable form for fitting to $D^{2}_{na}$ is $\tilde{g}(\lambda)$, which would suggest $D^{2}_{na}$ scales with $W^{p}$ at large strains.  
This form gives slightly less good fits to our data, but in any case the principle shown in Fig.\ \ref{fig:NAD} is the same as outlined in Ref.\ \cite{vorselaars09b}; for long chains, nearly affine deformation at large scales (in our language, $\lambda \simeq \lambda_{eff}$) increasingly drives nonaffine displacements (i.\ e.\ plastic activity) at smaller scales, leading to strain hardening.
This effect weakens and values for both $\sigma$ and $D^{2}_{na}$ decrease (for large strains) with decreasing $N$.

\section{Discussion and Conclusions}
\label{sec:discuss}

We derived a simple theory for polymeric strain hardening based on the notion that the increase in stress beyond yield amounts to an increase in the ``flow'' stress in an increasingly anisotropic viscoplastic medium.
In the strain hardening regime,  long and short chains relax on large scales via similar mechanisms ($\tau \propto \dot\epsilon^{-1}N$) and longer chains show greater hardening because they cannot relax on large length scales over the timescale of the deformation.  
In practice, this is due to the high interchain friction in the glassy state.

We theoretically predicted, and provided evidence using simulations, that coherent chain relaxation driven by resistance to chain contour length increase is a key factor in the large-strain mechanical response of uncrosslinked polymer glasses.
Coherent relaxation reduces the dominant (chain scale) relaxation time $\tau$ by a factor $N\dot\epsilon$ during active deformation.
Further, we claim that the increase in relaxation times when deformation is ceased is at least partially associated with the fact that relaxation need no longer be coherent.
Our results are consistent with many previous simulations and experiments, and semiquantitatively capture the increase in strain hardening as chain length increases.

Much recent work has emphasized the predominantly viscous/viscoelastic nature of polymeric strain hardening, at least prior to entanglement-stretching.
Recent experiments and modeling \cite{senden10,wendlandt10,hine07,buckley10} have provided strong support to the notion that entanglements play only a secondary role in glassy-polymeric strain hardening, at least for the majority of synthetic polymers and in the weak hardening regime.
Refs.\ \cite{hine07, buckley10} also emphasized the role of meltlike relaxation mechanisms in the marginally glassy state.

The present work, which focuses on relaxation mechanisms, is consistent with this these trends.
We quantitatively related strain hardening to local plasticity at the segmental scale, and showed
that the power-law dependence of the large-scale chain relaxation time $\tau$ on $N$ during active deformation, $\tau \propto N^{\gamma-1}$, is consistent with $\gamma = 2$, the same value as the Rouse model for unentangled polymer melts \cite{doi86,footz}.
Another ``Rouselike'' aspect is the apparent tight coupling of $\tau$ to a single microscopic relaxation time $\tau_{\alpha}$.

$\lambda_{eff}$ can now be accurately measured in scanning near-field optical microscopy (SNOFM) experiments \cite{ube07}, which have shown that $\epsilon_{eff} < \epsilon$ for entangled chains deformed slightly above $T_g$.
Analogous studies, well below $T_g$, could be performed to test the theory developed here; modern neutron scattering techniques might be employed for the same purpose.
Additionally, deformation calorimetry (DC) experiments \cite{adams88} can be performed to better understand the dissipative contribution to the stress.
We are not aware of any studies in which the SNOFM or DC methods have been applied to study strain hardening in polymer glasses.

Our model is minimal.  
It cannot quantitatively predict stress-strain curves either at small or at very large strains, because it neglects energetic components of stress and also strain softening, which play important roles in glassy polymer mechanics.
However, these limitations do not violate the spirit of our modeling effort, which was to illustrate the role of coherent relaxation in controlling large-scale chain conformations and influencing stress in actively deforming polymer glasses.

The theory presented here serves as a complement to a recent microscopic theory by Chen and Schweizer \cite{chen09}, which also provides a unified description of plastic flow and strain hardening in polymer glasses.
Ref.\ \cite{chen09} assumes $\bar{\lambda}_{eff}=\bar{\lambda}$, and thus appears to neglect an important relaxation mechanism.
Also, Ref.\ \cite{chen09} is based on an extension of liquid state theory to the glassy state \cite{chen07} and (formally) breaks down at zero temperature.
In contrast, the theory developed here assumes relaxation is dominated by strain activated processes, and so should be most accurate at low temperatures; we expect it to break down concurrently with the validity of the ``mean field'', independent-chain-relaxation behavior \cite{hoy09b} of $\lambda_{eff}$ as $T\to T_g$.

An ideal approach would account for both thermally and strain activated relaxation without sacrificing simplicity.
For example, $\tau$ should have explicit temperature dependence, and thermal activation should produce logarithmic corrections to $\dot\epsilon^{-1}$ scaling.
The dependence of $\tau$ on $|\sigma|$ (from stress-assisted rearrangements) should also be captured. 
Combining our theory with a more sophisticated theory for $\sigma_{flow}$ (e.\ g.\ Refs.\  \cite{chen07} or \cite{boyce88}) might be fruitful.
Alternatively, modern techniques in nonequilibrium thermodynamics may prove a useful tool for treating finite-$T$ effects.

Clearly, further work is necessary to quantitatively predict stress-strain and stress-relaxation curves.
In particular, improved understanding of the effects of chain stiffness is required. 
It seems certain that microscopic structural detail at the Kuhn scale (e.\ g.\ chemistry-dependent effects) exerts significant influence on segmental relaxation processes, and it is probable that these effects couple to chain-scale relaxation (see e.\ g.\ Refs.\ \cite{capaldi02, lyulin06, govaert08, hintermeyer08, vorselaars09b}).
Studies using more chemically realistic models, or real polymers, would be welcome.

All MD simulations were performed using LAMMPS \cite{plimpton95}.  
Mark O.\ Robbins contributed significantly to this project, with numerous helpful discussions. 
Kenneth S.\ Schweizer,  Grigori Medvedev, Kang Chen, Daniel J.\ Read, and Edward J.\ Kramer also provided helpful discussions, and  K.\ S.\ provided the original concept for $\lambda_{eff}$.
Gary S.\ Grest provided equilibrated $N=500$ states.
The authors wish to thank the KITP Glasses '10 conference, which was the setting of many of these discussions.
Support from NSF Awards No.\ DMR-0520415 (RH) and DMR-0835742 (RH, CO) is gratefully acknowledged.


\begin{thebibliography}{68}
\bibitem[{foo({\natexlab{h}})}]{footd}
\bibinfo{note}{A more accurate form for this increase which accounts for
  non-Gaussian effects predicts $\sigma \sim g^{*} = \ell^{-1}(\tilde{g}(\lambda/\sqrt{N_e})g(\lambda)$, where $\ell$ is the
  Langevin function \cite{arruda93b}.}
\bibitem[{foo({\natexlab{a}})}]{footi}
\bibinfo{note}{Very short chains do not significantly orient, and (if brittle
  fracture does not intervene) oligomeric glasses show nearly perfect-plastic
  flow, similar to that observed in atomic glasses \cite{hoy06,hoy07};
  Eq.\ \ref{eq:sigmaeff} still applies in this case because $\bar\lambda_{eff}
  \simeq \bar{I}$ and $g(\bar{I}) = 0$.}
\bibitem[{foo({\natexlab{c}})}]{foot5}
\bibinfo{note}{Restoration of $\epsilon_{eff}$ towards its equilibrium value
  (zero) can in principle, drive shape recovery in a stress relaxation experiment; motions decreasing $|\epsilon_{eff}|$ correspond roughly to the
  ``reverse shear transformations'' \cite{argon73} studied by many authors. However, treatment
  of such effects is beyond the scope of this work. In practice, the time scale
  of relaxation of $|\epsilon_{eff}|$ after a large deformation would be very
  long if $T$ is well below $T_g$.}
\bibitem[{foo({\natexlab{d}})}]{footq}
\bibinfo{note}{In practice, systems are ``densely'' entangled if $N_e l_0
  \lesssim 10l_K$ \cite{hoy07}.}
\bibitem[{foo({\natexlab{e}})}]{foot1}
\bibinfo{note}{Eq.\ \ref{eq:epseffoft} predicts ``saturation'' of orientation
  and hardening at large strains $\epsilon_{eff} \gtrsim \epsilon_{eff}^{sat} =
  \dot{\epsilon}\tau$; it can be rewritten as $\epsilon_{eff}(\epsilon) =
  \epsilon_{eff}^{sat} \left(1-\exp(-\epsilon/\epsilon_{eff}^{sat})\right)$. An
  interesting property of this result is that since $\tau$ increases
  monotonically with $N$, so will $\epsilon_{c}^{sat}$. For long chains, our
  theory will break down in the limit where chains are pulled taut,
  $\tilde{g}(\epsilon_{eff}^{sat}) \sim Nl_0/l_K$.  However, this limit is
  difficult to reach at experimentally accessible strains.}
\bibitem[{foo({\natexlab{y}})}]{footy}
\bibinfo{note}{In Section \ref{subsec:stresstheory}, $u_0$ and $a$ need not necessarily correspond to the Lennard-Jones interactions employed in our simulations.  They are intended to be interpreted more generally.}
\bibitem[{foo({\natexlab{g}})}]{footl}
\bibinfo{note}{Other recent work \cite{wendlandt05,senden10,wendlandt10} relates the rate dependence and nonlinearity of strain hardening to an Eyring-like model in which the activation volume $V$ decreases with increasing strain.  In contrast, we predict that $V$ increases. An important difference between the approaches is that in Eyring-like models, $V$ is a `parameter' with the dimensions of volume.}
\bibitem[{foo({\natexlab{k}})}]{footp}
\bibinfo{note}{This asymmetry has been treated in other work (see e.\ g.\
  Refs.\ \cite{boyce88,rottler03c}) and could in principle be treated by
  modifying $W^s$ appropriately.}
\bibitem[{foo({\natexlab{l}})}]{foot2}
\bibinfo{note}{Interestingly, however, our theoretical stress strain curves are
  remarkably similar to those in preoriented systems where deformation is
  applied along the preorientation direction; see, e. g., Fig.\ 1(b) of Ref.\
  \cite{ge10} and Fig.\ 3(b) of Ref.\ \cite{senden10}.}
\bibitem[{foo({\natexlab{m}})}]{footz}
\bibinfo{note}{Our results do not exclude the possibility of a crossover to higher $\gamma$ for chains with $N \gg N_e$. 
Note that experimental data for quiescent dynamics near $T_g$ \cite{hintermeyer08} suggests that any such crossover may be chemistry-dependent.
While our theory explicitly neglects entanglement effects, these might be captured by adjusting $\gamma$ appropriately.
Few systematic experimental studies of the variation of glassy polymer mechanical response with $N$ have been published, largely because of the difficulty of avoiding fracture at small $N$. 
Ref.\ \cite{govaert04} presented data for $N \gg N_e$ systems that suggests hardening and melt relaxation vary similarly with chain length (i.\ e., both are dominated by large-scale relaxation).
Although data was interpreted in terms of relaxation of the entanglement network, the results are not inconsistent with ours because $\gamma$ was not measured.}
\end{thebibliography}

\end{document}